RESEARCH

# Pyk2 plays a critical role in synaptic dysfunction during the early stages of Alzheimer's disease

Quentin Rodriguez[1*], Floriane Payet[1], Karina Vargas-Baron[1], Eve Borel[1], Fabien Lanté[1], Sylvie Boisseau[1], Béatrice Blot[1], Jean-Antoine Girault[2,3] and Alain Buisson[1*]

## Abstract

**Background** The locus of the gene *PTK2B* encoding the tyrosine kinase Pyk2 has been associated with the risk of late-onset Alzheimer's disease, the predominant form of dementia. Pyk2 is primarily expressed in neurons where it is involved in excitatory neurotransmission and synaptic functions. Although previous studies have implicated Pyk2 in amyloid-β and Tau pathologies of Alzheimer's disease, its exact role remains unresolved, with evidence showing both detrimental and protective effects in mouse models. Here, we investigate the role of Pyk2 in hippocampal hyperactivity, Tau synaptic localization and synaptic loss associated with Alzheimer's disease-related alterations occurring in the early stages of the disease.

**Methods** Pyk2's involvement in amyloid-β oligomer-induced hippocampal neuronal hyperactivity was investigated using whole-cell patch clamp in hippocampal slices from WT and Pyk2 KO mice. Various Pyk2 mutants were overexpressed in cultured cortical neurons to study Pyk2's role in synaptic loss. Pyk2 and Tau interaction was assessed with bimolecular fluorescence complementation assays in cultured neurons and co-immunoprecipitation in mouse cortex. To evaluate the impact of Pyk2 on Tau expression in synapses, cellular fractionation was performed on hippocampi from WT and Pyk2 KO mice.

**Results** Genetic deletion of Pyk2 prevented amyloid-β oligomer-induced hippocampal neuronal hyperactivity and synaptic loss. Overexpression of Pyk2 in neurons decreased dendritic spine density independently of its autophosphorylation or kinase activity, but through its proline-rich motif 1. Furthermore, Pyk2 interacted with Tau in synapses, while Pyk2 deletion decreased Tau synaptic localization in the hippocampus.

**Conclusions** Pyk2 contributes to hippocampal neuronal hyperactivity and synaptic loss, two early events in Alzheimer's disease pathogenesis. It is also involved in Tau synaptic localization, a process known to be detrimental in Alzheimer's disease. These findings highlight Pyk2 as a critical player in Alzheimer's disease pathophysiology and suggest its potential as a promising therapeutic target for early intervention.

**Keywords** Alzheimer's disease, Hippocampal neuronal hyperactivity, Synaptic loss, Pyk2, Aβ oligomers, Tau

*Correspondence:
Quentin Rodriguez
quentin.rodriguez@univ-grenoble-alpes.fr
Alain Buisson
alain.buisson@univ-grenoble-alpes.fr
Full list of author information is available at the end of the article

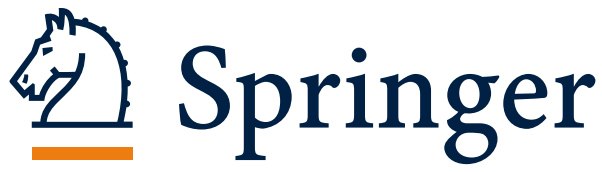

## Background

Alzheimer's disease (AD), the most common form of dementia, affects approximately 32 million people worldwide [1]. Projections suggest that the number of patients will nearly double every 20 years [2]. AD is a progressive disease that begins 10 to 20 years before the onset of clinical symptoms, transitioning from a preclinical stage through mild cognitive impairment (MCI) to dementia.

Late-onset AD, the most frequent form of the disease, results from a combination of environmental and genetic risk factors, with the genetic component accounting for 60–80% of the risk [3]. Genome-wide association studies have associated the *PTK2B* locus, encoding the tyrosine kinase Pyk2, with the risk of late-onset AD [4–8].

Pyk2 has been implicated in the pathophysiology of AD, particularly in amyloid-β (Aβ)-induced synaptotoxicity, though its role remains controversial. Studies have demonstrated both deleterious and protective effects of Pyk2 in AD mouse models. Genetic deletion or pharmacological inhibition of Pyk2 in APPswe/PS1ΔE9 mice prevents synaptic plasticity disruption, synaptic loss, and improves behavioral deficits induced by Aβ oligomers (Aβo) [9–11]. Conversely, Pyk2 overexpression in the hippocampus of 5xFAD mice increases synaptic density and decreases memory deficits that characterize this AD mouse model [12].

If the majority of the studies have focused on Pyk2's role in Aβ-related pathophysiology, fewer have explored the influence of Pyk2 on Tau pathology in AD. Emerging evidence suggests a role for Pyk2 in Tau phosphorylation. Indeed, Pyk2 has been shown to colocalize with hyperphosphorylated Tau and Tau oligomers in the brain of Alzheimer's patients and in the brain of a tauopathic mouse model [13]. *In vitro* studies have demonstrated that Pyk2 interacts with Tau and phosphorylates its tyrosine 18, while Pyk2 overexpression increases Tau phosphorylation in P301L Tau transgenic mice [14]. Additionally, Pyk2 activates GSK-3β, a kinase known to phosphorylate Tau [15–17], suggesting that Pyk2 influences Tau phosphorylation levels. However, recent findings have described a suppressive effect of Pyk2 on Tau pathology [18]. Specifically, genetic deletion of Pyk2 in PS19 mice, a tauopathic mouse model, resulted in increased Tau phosphorylation and accumulation in the brain, decreased neuronal survival, and impaired spatial memory. These results challenge the idea of a deleterious role for Pyk2 in Tau pathology.

Concomitantly with the early accumulation of Aβ and phosphorylated Tau in the brain of AD patients, two pathological features have been identified during the preclinical and/or MCI stages of AD: hippocampal and cortical hyperactivity [19–22] and synaptic loss [23]. Our previous work demonstrated hippocampal neuronal hyperactivity in APP/PS1-21 (APP/PS1) mice as one of the first pathophysiological markers of the disease [24]. Given the involvement of Pyk2 in excitatory neurotransmission [25–27], we investigated its role in neuronal hyperactivity. We found that genetic deletion of Pyk2 prevents Aβo-induced neuronal hyperactivity and synaptic loss. While neuronal hyperactivity correlates with increased Pyk2 phosphorylation in the postsynaptic compartment of neurons, synaptic loss is not influenced by Pyk2 activity but involves its proline-rich motif 1 (PR1). When we explored Pyk2 and Tau relationship, we demonstrated their interaction in dendritic spines and showed that genetic deletion of Pyk2 decreases Tau expression in the postsynaptic compartment. Together, these findings implicate Pyk2 in the early stages of AD, particularly in Aβo-induced neuronal hyperactivity and synaptic loss, and suggest its role in pathological Tau accumulation in synapses.

## Methods

### Animals

All experiments were conducted in accordance with the European Community Council directive 86/609/EEC (November 24, 1986) and French national institutional animal care guidelines (protocol APAFIS#45114). All experimental protocols were approved by the Grenoble Institute of Neurosciences Ethics Committee. Three transgenic mouse models on a C57BL/6J genetic background were used in this study. APP/PS1-21 mice co-express the human Swedish mutation (KM670/671NL) of the amyloid precursor protein (APPswe) and the human L166P mutation of the Presenilin-1 (PS1), both under *Thy1* promoter control [28]. Pyk2 KO mice were generated as previously described [29]. Briefly, homologous recombination in embryonic stem cells was performed to delete exons 15 to 18, which correspond to the kinase domain. Splicing between exons 14 and 19 creates a premature STOP codon in exon 19. Pyk2 KO mice were genotyped using tail biopsies with the following primers: Pyk2 forward: 5'-TGTGCTCAGAGAAAAACG GAGGAACCCT-3', Pyk2 reverse 1: 5'-CATTGATTC CTGCTTCAGCCCTGGTCTAA-3' and Pyk2 reverse 2: 5'-GCCCATCGGGGCGATTTAAATATAATTCG-3'. Tau KO mice, generously provided by Dr. Isabelle Arnal (Grenoble Institute of Neurosciences), were generated as previously described [30]. All experiments were performed with no preference for either female or male animals. The mice were housed in groups at the Grenoble Institute of Neurosciences, with a 12-h light/dark cycle and *ad libitum* access to food and water.

#### Primary culture of cortical neurons

Cortical neurons were obtained from embryonic day 15 ± 1 Swiss CD1 or C57BL/6J mice. Briefly, cerebral cortices were dissected, mechanically dissociated, and cultured in Dulbecco's Modified Eagle's Medium (DMEM) – high glucose (Sigma-Aldrich #D5671) supplemented with 5% horse serum (Gibco #16050122), 5% fetal bovine serum (Sigma-Aldrich #F7524), and 2 mM L-glutamine (Gibco #A2916801). For confocal microscopy, neurons were cultured in 35 mm glass-bottom dishes (MatTek #P35G-0–14-C). 12-well plates (NEST #712001) were used for biochemical experiments. Dishes and plates were pre-coated with 0.1 mg/mL poly-D-lysine (Sigma-Aldrich #P6407) and 0.02 mg/mL laminin (Sigma-Aldrich #L2020). Cultures were incubated at 37°C in a humidified atmosphere of 5% CO2/95% air until used for experiments. For biochemical experiments, neurons in 12-well plates were washed with DMEM and treated with 100 nM Aβo for 1, 6 (at 14 DIV), or 24 h (at 13 DIV).

#### Plasmids

Plasmids encoding rat GFP-Pyk2 WT, GFP-Pyk2 Y402F, GFP-Pyk2 K457A, and GFP-$Pyk2_{421-1009}$ were generated as previously described [31, 32]. Rat Pyk2 shares 98% amino acid identity with mouse Pyk2, with strong conservation of functional domains. The Pyk2 cDNAs of these constructs were cloned into the pmCherry-C1 vector (Clontech) using SacI and KpnI restriction enzymes to generate mCherry-Pyk2 constructs. The rat Pyk2 (1–368) sequence was amplified by PCR from the mCherry-Pyk2 WT construct and cloned into the pmCherry-C1 vector (SacI/KpnI) to generate the mCherry-$Pyk2_{1-368}$ plasmid. The plasmid encoding mCherry-Pyk2 P413/416A was generated by site-directed mutagenesis of the mCherry-Pyk2 WT plasmid. Plasmids encoding LifeAct-GFP and LifeAct-RFP were obtained from Ibidi (vectors pCMV-LifeAct-TagGFP2 and pCMV-LifeAct-TagRFP, respectively). LifeAct is a 17-amino-acid peptide that specifically binds to filamentous actin, thereby allowing the visualization of dendritic spines [33]. cDNAs of mouse BACE1 and rat Pyk2 WT were cloned into the pBiFC-VN173 vector (Addgene) using SacI and HindIII restriction enzymes. Human Tau cDNA was cloned into the pBiFC-VC155 vector (Addgene) using SalI and BglII restriction enzymes. The BACE1-VC155 construct was generated by replacing GFP in the peGFP vector (Clontech) with the mouse BACE1-VC155 sequence using PstI and NotI restriction enzymes. The human APPswe-GFP-GSG-T2A-LifeAct-GFP construct was obtained by inserting the human APPswe-GFP-GSG-T2A sequence into the pCMV-LifeAct-TagGFP2 plasmid, upstream of the LifeAct-GFP sequence, using NheI and XhoI restriction enzymes. The GSG-T2A sequence (referred to as T2A) enables the production of two distinct proteins from a single plasmid through a ribosomal skipping mechanism occurring between the glycine and proline residues at the C-terminal end of the T2A sequence [34]. All plasmids were constructed using the Pro Ligation-Free Cloning Kit (Applied Biological Materials #E087) and verified by sequencing. Plasmid purification was performed using the NucleoBond Xtra Midi EF kit (MACHEREY–NAGEL #740420.50) according to the manufacturer's protocol.

#### Transfection

Neuronal transfections were performed in cortical neuron cultures at 11–13 DIV using calcium phosphate precipitation. The growth medium (DMEM supplemented with sera) was removed and maintained at 37°C for later use. Cells were washed with DMEM and incubated at 37°C for 30 min in DMKY buffer (1 mM kynurenic acid, 0.9 mM NaOH, 0.5 mM HEPES, 10 mM $MgCl_2$, 0.05% phenol red, pH 7.4). Concurrently, 3 µg of each plasmid was mixed with 120 mM $CaCl_2$ and HEPES-buffered saline (25 mM HEPES, 140 mM NaCl, 0.75 mM $Na_2HPO_4$, pH 7.06) and incubated for 15 min at room temperature to precipitate the DNA. This mixture was applied to neurons for 2 h. Subsequently, neurons were washed with DMEM, which was then replaced by the previously removed growth medium maintained at 37°C. Neurons were returned to the incubator for 48 h before confocal microscopy imaging (13 to 15 DIV).

To verify Pyk2 expression levels of the different mutants (mCherry-Pyk2 WT, Y402F, K457A, P413/416A) and truncated forms (mCherry-Pyk2 1–368 and 421–1009), HEK293 cells were transfected using jetPEI reagent (Ozyme # POL101000053) according to the manufacturer's protocol. Cells were transfected with 3 µg of each plasmid and were lysed 48 h post-transfection for immunoblot analysis.

#### Confocal imaging and spine density analysis

The medium of transfected neurons was replaced with Hanks' Balanced Salt Solution containing (in mM): 110 NaCl, 5 KCl, 2 $CaCl_2$, 0.8 $MgSO_4$, 1 $NaH_2PO_4$, 12 HEPES, 5 D-glucose, 25 $NaHCO_3$, 0.01 glycine. Neurons were visualized using a Nikon Ti C2 confocal microscope equipped with a Nikon 60X water immersion objective and NIS-Elements software (Nikon, Melville, New York, USA). GFP and mCherry/RFP fluorophores were excited at 488 nm (emission filtered at 504–541 nm) and 543 nm (emission filtered at 585–610 nm), respectively. Images were acquired as Z-stacks with a step size of 0.3 µm. The acquired images were then deconvolved using AutoQuantX3 software (Media Cybernetics, Abingdon, Oxon, UK). Spine density was determined using NeuronStudio software (Icahn School of Medicine at Mount Sinai, New York, USA), which automatically measured dendrite length and the number of associated spines. Images were scaled using ImageJ (v1.54f). Bilinear interpolation was used to increase image size while minimizing the introduction of artifacts. This scaling was applied solely to improve the visual presentation of the images and was not used for any quantitative analysis.

### Production and purification of Aβ-poly-histidine oligomers

The cDNA encoding human $A\beta_{1-42}$ peptide with a 6xHis-tag was cloned into the pet28a vector (Novagene) using NdeI and PspXI restriction enzymes. The plasmid was then transformed into *Escherichia coli* (LGC Biosearch Technologies #60107–2), followed by overnight incubation in Luria–Bertani medium at 37°C. Protein expression was induced with 1 mM isopropyl β-D-1-thiogalactopyranoside for 4 h at 37°C, after which bacteria were harvested by centrifugation at 5,500 × g for 20 min at 4°C. The pellet was resuspended in ice-cold phosphate-buffered saline (PBS, 5 mL per gram of pellet) supplemented with protease (Sigma-Aldrich #P8340-5ML) and phosphatase inhibitors (Sigma-Aldrich #P5726-5ML) (1:50). Cells were lysed on ice by sonication (7 cycles of 1 min ON and 30 s OFF at 40% amplitude) and centrifuged at 15,000 × g for 10 min at 4°C. The cytosolic supernatant was collected, while the pellet was resuspended in 10 mL of PBS containing 8 M urea, sonicated again and centrifuged as before. The resulting supernatant, containing inclusion bodies enriched in Aβ peptides, was combined with the cytosolic supernatant and filtered (0.45 µm pore size) to remove cellular debris. The clarified supernatant (inclusion bodies + cytosol) was diluted 1:1 in binding buffer (PBS, 10 mM imidazole, pH 7.4) with protease and phosphatase inhibitors (1:50) and incubated on Ni–NTA agarose resin (MACHEREY–NAGEL) for 1 h at 4°C with agitation. The mixture was loaded onto the column, washed with washing buffer (PBS, 20 mM imidazole, pH 7.4), and His-tagged Aβ peptides were eluted using elution buffer (PBS, 500 mM imidazole, pH 7.4). The concentration of Aβ was determined using a bovine serum albumin (BSA) standard curve via SDS-PAGE analysis. The purified protein was stored at −20°C.

### Subcellular fractionation

To assess protein expression at the synaptic level, subcellular fractionations were conducted as previously described [35]. Cultured neurons or mouse hippocampi were homogenized in solution 1 (320 mM sucrose, 10 mM HEPES, pH 7.4) and centrifuged at 1,000 × g for 10 min to remove nuclei and debris. The resulting supernatant was centrifuged at 12,000 × g for 20 min to obtain a crude membrane fraction. The pellet was resuspended in solution 2 (4 mM HEPES, 1 mM EDTA, pH 7.4) and centrifuged twice at 12,000 × g for 20 min. The pellet was then reconstituted in solution 3 (20 mM HEPES, 100 mM NaCl, 0.5% Triton X-100 for cells and 1% Triton X-100 for tissues, pH 7.2), incubated for 1 h at 4°C with agitation, and centrifuged at 12,000 × g for 20 min. The supernatant was collected as the non-postsynaptic density (non-PSD) fraction (Triton-soluble). The remaining pellet was resuspended in solution 4 (20 mM HEPES, 150 mM NaCl, 1% Triton X-100, 1% deoxycholic acid, 1% SDS, pH 7.5), incubated for 1 h at 4°C with agitation, and then centrifuged at 10,000 × g for 15 min. The resulting supernatant was collected as the PSD fraction (Triton-insoluble). Samples were maintained at 4°C during all steps of the experiment. Protease and phosphatase inhibitors (1:100) were added to all solutions immediately before use. The integrity of the PSD and non-PSD fractions was confirmed by immunoblotting using synaptophysin and PSD95 antibodies, which are respectively enriched in the non-PSD and PSD fractions.

### Immunoblotting

Samples (cells or tissues) were mechanically lysed on ice in RIPA buffer (50 mM Tris–HCl, pH 8, 150 mM NaCl, 1% NP-40, 0.5% sodium deoxycholate, 0.1% SDS) supplemented with protease and phosphatase inhibitors (1:100). Protein concentration was determined using the Pierce BCA assay kit (Thermo Scientific #23227) and measured with a Pherastar plate reader (BMG Labtech). The samples were subsequently diluted in Laemmli buffer (Bio-Rad #1610747) containing 10% β-mercaptoethanol and heated at 95°C for 10 min. SDS-PAGE was performed using an equal amount of protein (10 to 20 µg) on 10% polyacrylamide pre-cast gels (Bio-Rad #4568034). Proteins were transferred to a polyvinylidene fluoride membrane (Bio-Rad #1704156) using the Transblot Turbo system (Bio-Rad). Following transfer, membranes were blocked in Tris-buffered saline containing 0.1% Tween 20 (TBS-T) and 5% BSA for 2 h at 37°C. Membranes were then incubated overnight at 4°C with the following primary antibodies: anti-total Pyk2 (Cell Signaling #3480, 1:2000), anti-pPyk2 Tyr402 (Invitrogen #BS-3400R, 1:1000), anti-tubulin β3 (BioLegend #801201, 1:10,000), anti-PSD95 (Merck Millipore #MABN68, 1:10,000), anti-synaptophysin (Merck Millipore #MAB329-C, 1:1000), anti-total Tau (Dako #A0024, 1:10,000), anti-RFP (Rockland #600–401-379, 1:1000). After successive washes in TBS-T, membranes were incubated with HRP-conjugated secondary antibodies (Jackson ImmunoResearch; 1:40,000) for 1 h at room temperature. Specific proteins were revealed using Immobilon ECL Ultra solution (Sigma-Aldrich #WBULS0500) and detected with the Chemidoc detection system (Bio-Rad). Analysis of the results was conducted using ImageJ or Image Lab software.

### Co-immunoprecipitation

Cortical tissues from 3-month-old WT and Tau KO mice were mechanically lysed in RIPA buffer supplemented with protease and phosphatase inhibitors (1:100). The lysate was then centrifuged at 1,000×g for 10 min at 4°C, and the protein concentration of the resulting supernatant (input) was determined using the Pierce BCA assay kit. Next, 250 µg of lysate was precleared twice with 20 µL of magnetic beads coupled to protein A (Cytiva #28944006) for 24h at 4°C. Protease and phosphatase inhibitors (1:100) were added between the two preclearing steps. Subsequently, 20 µL of magnetic beads coupled to protein A were washed with TBS (50 mM Tris, 150 mM NaCl, pH 7.5) and incubated for 1 h at room temperature on a rotating wheel with anti-total Tau antibody (Dako #A0024, 1:50) diluted in TBS with 5% BSA. The precleared lysate was then added to the bead–antibody complexes and incubated 15 min at room temperature on a rotating wheel. The immunoprecipitated complexes were washed three times with RIPA buffer, and the proteins bound to the beads were eluted with elution buffer (PBS, Laemmli buffer containing 10% β-mercaptoethanol). The eluate was heated at 95°C for 10 min, and immunoblotting was performed to identify the co-immunoprecipitated proteins.

### Preparation of human Aβo for electrophysiology experiments

Aβo were prepared from recombinant human $A\beta_{1-42}$ peptide (Bachem #4014447). The lyophilized peptide was initially dissolved in 1,1,1,3,3,3-hexafluoro-2-propanol (HFIP) to a concentration of 1 mM as previously described [36]. After HFIP evaporation, the resulting peptide film was resuspended to 100 µM in dimethyl sulfoxide and then diluted in ice-cold artificial cerebrospinal fluid (aCSF). This solution was immediately vortexed and sonicated for 1 h at 4°C, followed by continuous vortexing at 4°C for 24 h to promote Aβ oligomer formation.

### Hippocampal slice preparation

Hippocampal slices were prepared from 25 to 32-day-old mice (male and female). Mice were euthanized by cervical dislocation and decapitated. The brain was rapidly removed and sagittal slices (300 µm thickness) were cut using a vibratome VT1200S (Leica) in ice-cold cutting solution containing (in mM): 2.5 KCl, 1.25 $NaH_2PO_4$, 10 $MgSO_4$, 0.5 $CaCl_2$, 26 $NaHCO_3$, 234 sucrose, 11 D-glucose, saturated with 95% $O_2$ and 5% $CO_2$. The hippocampus was then micro-dissected and incubated in aCSF containing (in mM): 119 NaCl, 2.5 KCl, 1.25 $NaH_2PO_4$, 1.3 $MgSO_4$, 2.5 $CaCl_2$, 26 $NaHCO_3$, 11 D-glucose, saturated with 95% $O_2$ and 5% $CO_2$, at 37°C for 30 min. Subsequently, slices were maintained at room temperature for at least 1 h before recording. For biochemical experiments, hippocampal slices were treated with 100 nM Aβo for 20 min before lysis.

### Whole-cell patch clamp recordings

Hippocampal slices from male and female mice were transferred to a submerged recording chamber continuously perfused with oxygenated aCSF (2 mL/min) at room temperature. Whole-cell recordings were made in the soma of CA1 pyramidal neurons visualized using a DIC camera on an upright microscope (Nikon Eclipse E600FN). Borosilicate glass pipettes (5–6 MΩ) were filled with an intracellular solution containing (in mM): 117.5 $CsMeSO_4$, 15.5 CsCl, 10 TEACl, 8 NaCl, 10 HEPES, 0.25 EGTA, 4 MgATP, 0.3 NaGTP, 5 QX-314 (290 mOsm, pH 7.3). Spontaneous excitatory postsynaptic currents (sEPSCs) were recorded at a membrane holding potential of −60 mV in the presence of picrotoxin (5 µM), a $GABA_A$ receptor antagonist, to isolate excitatory currents. Signals were acquired using a double EPC 10 amplifier (HEKA Elektronik), filtered at 1 kHz, and sampled at 10 kHz. Currents were acquired with PatchMaster software (HEKA Elektronik) and analyzed using Mini Analysis Program software. Input and access resistance were constantly monitored, and neurons showing > 20% changes in these parameters were excluded. A threshold amplitude was set to 5 pA, and all detected events were accepted or rejected based on visual examination.

For experiments with APP/PS1 mice, sEPSCs were recorded for 5 min after a 5-min stabilization period. For experiments with Aβo, currents were recorded for 5 min (baseline) after a 5-min stabilization period, followed by 20 min of recording with 100 nM Aβo perfusion. For each neuron, sEPSC changes were normalized to the baseline using the last 5 min of each recording.

### Statistical analysis

Data were analyzed using GraphPad Prism 8.4 software. For statistical comparisons between two groups, the two-tailed Mann–Whitney test was applied. Multiple comparisons were performed using the Kruskal–Wallis test followed by Dunn's post-hoc test for non-normally distributed data, while a one-way ANOVA test followed by Tukey's post-hoc test was applied for normally distributed data. Results are expressed as mean ± SEM from independent biological samples. Sample sizes for each dataset are indicated in the figure legends.

## Results

### Pyk2 mediates Aβo-induced hippocampal neuronal hyperactivity

In Alzheimer's patients, hippocampal hyperactivity has been identified in the early stages of the disease (MCI stage), followed by a reduction in brain activity in later stages [20]. More recently, early cortical hyperactivity has been found in asymptomatic adults who later developed AD. In this study, hyperactivity has been associated with Aβ deposition [22]. Here, we used patch clamp recordings in hippocampal slices and found that the APP/PS1-21 mouse model of AD exhibited neuronal hyperactivity at 1 month of age, before any structural or cognitive defects, as indicated by a 1.8-fold increase in sEPSCs frequency and a 1.4-fold increase in amplitude compared to WT mice (Fig. 1A and B). This hippocampal hyperactivity can be reproduced by Aβo exposure in acute hippocampal slices from WT animals [24, 37]. To decipher Pyk2's involvement in early neurotransmission alterations in AD models, we recorded sEPSCs from CA1 pyramidal neurons using whole-cell patch clamp in hippocampal slices from 1-month-old WT or Pyk2 KO mice, both incubated or not with Aβo. We analysed the neuronal excitatory currents using the $T_{20}/T_0$ ratio, $T_0$ corresponds to spontaneous excitatory activity recorded for 5 min prior to treatment with 100 nM Aβo or vehicle, and $T_{20}$ represents the spontaneous activity recorded during the last 5 min of a 20-min treatment. In WT mice treated with Aβo, we observed a 42% and 18% increase in sEPSCs frequency and amplitude, respectively (Fig. 1C and D). In contrast, Aβo exposure did not change sEPSC frequency and amplitude in Pyk2 KO mice (Fig. 1E and F). The genetic deletion of Pyk2 prevents Aβo-induced hyperactivity in hippocampal neurons, suggesting a role for Pyk2 in mediating neuronal hyperactivity.

### Neuronal hyperactivity is associated with Pyk2 phosphorylation at Tyr402

Studies have revealed that Aβo-induced hyperactivity is promoted by glutamate accumulation in the synaptic cleft, resulting in the overactivation of ionotropic glutamatergic receptors and the subsequent increase in intracellular calcium levels [38, 39]. In this context, since Pyk2 is a calcium-dependent tyrosine kinase that requires trans-autophosphorylation at Tyr402 in order to be active [40–42], we assessed the phosphorylation status of Pyk2 located in dendritic spines when neuronal hyperactivity occurs in hippocampal neurons from 1-month-old APP/PS1 mice and WT mice perfused for 20 min with 100 nM Aβo. We performed cellular fractionation on hippocampal slices from these mice to isolate the PSD fraction and evaluate the phosphorylation state of Pyk2 (Fig. 2A-D). We observed a 1.4-fold increase in Pyk2 phosphorylation at Tyr402 in the PSD fraction of both APP/PS1 mice and WT mice treated with 100 nM Aβo compared to untreated WT mice, with no change in total amount of Pyk2. We also observed a 1.3-fold increase in total amount of Tau in the PSD fraction. We also examined Pyk2 phosphorylation in older mice and observed a significant increase in 2-month-old APP/PS1 mice compared to WT, but no changes at 3 or 6 months when APP/PS1 neurons become hypoactive (**Supplementary Fig. 1**). Hence, when hippocampal neurons displayed an Aβo-driven hyperactivity, Pyk2 phosphorylation at Tyr402 and total Tau are concomitantly increased.

### Pyk2 interacts with Tau in neurons

We showed above the implication of Pyk2 in Aβo-induced neuronal hyperactivity, which coincided with an increase in phosphorylated Pyk2 and Tau content in the postsynaptic compartment. Furthermore, other studies have established a connection between Pyk2 and Tau [13, 14, 18]. Consequently, we focused our investigation on the relationship between these two proteins.

To elucidate the link between Pyk2 and Tau, we first conducted co-immunoprecipitation experiments using cortical lysates from 3-month-old WT mice, with age-matched Tau KO mice as negative control. Immunoprecipitation of Tau resulted in co-immunoprecipitation of Pyk2 (Fig. 3A). These findings suggest an interaction between Pyk2 and Tau. To further corroborate this interaction and determine its subcellular localization, we performed bimolecular fluorescence complementation (BiFC) assays. Cultured neurons were transfected with plasmids encoding Pyk2 WT fused to one half of GFP (VN) and Tau 1N4R fused to the complementary half (VC). The underlying principle of this technique is that protein–protein interaction brings the two GFP halves into proximity, resulting in a green fluorescence emission (Fig. 3B). Following the transfection with LifeAct-RFP and plasmids encoding Pyk2 VN and Tau VC, we observed green fluorescence throughout the neurons, including the soma, neurites and dendritic spines (Fig. 3C). This observation indicates an interaction between Pyk2 and Tau in neurons. To confirm the specificity of this interaction and rule out the possibility of random associations between the GFP halves, we conducted control experiments using plasmids encoding beta-site APP cleaving enzyme 1 VN (BACE1 VN) or BACE1 VC (**Supplementary Fig. 2**). Neurons overexpressing either Pyk2 VN and BACE1 VC or BACE1 VN and Tau VC did not exhibit any green fluorescence,

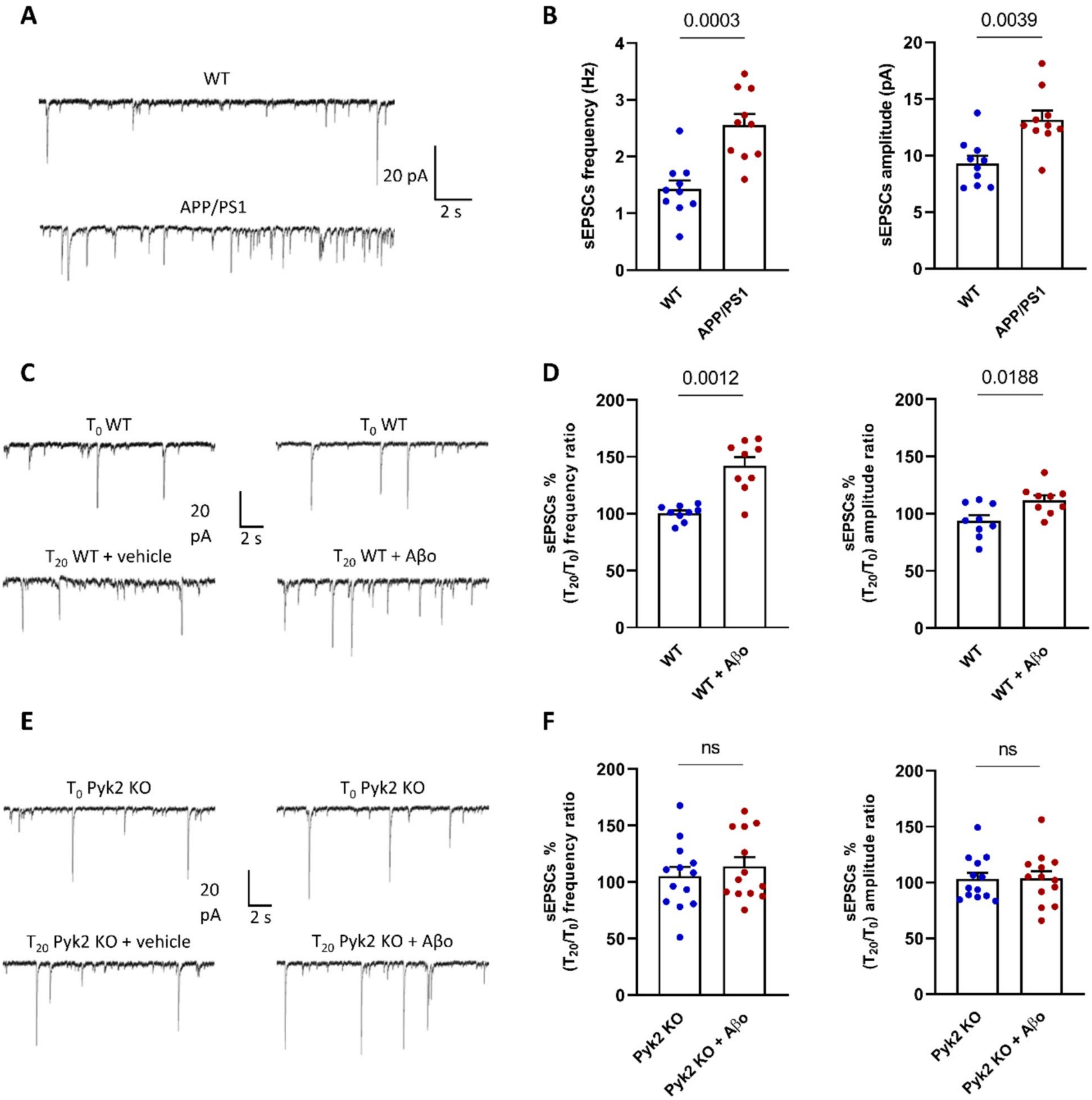

**Fig. 1** Pyk2 is involved in Aβo-induced hyperactivity of hippocampal neurons in 1-month-old mice. **A** Representative traces of sEPSCs recorded in whole-cell mode from hippocampal neurons clamped at −60 mV in 1-month-old WT and APP/PS1-21 mice. **B** Bar graphs showing the frequency and amplitude of sEPSCs recorded from pyramidal neurons in the CA1 region of the hippocampus in 1-month-old WT and APP/PS1-21 mice ($n = 10$ neurons from 3 WT mice with 3 to 4 neurons per mouse; $n = 10$ neurons from 3 APP/PS1-21 mice with 3 to 4 neurons per mouse). Mann–Whitney tests. **C, E** Representative traces of sEPSCs recorded in whole-cell configuration from hippocampal neurons clamped at −60 mV in 1-month-old WT (**C**) and Pyk2 KO (**E**) mice. Traces were extracted at the beginning of the experiment (T0) and 20 min later (T20), either in the presence of vehicle or Aβo, left and right panels, respectively. **D, F** Bar graphs showing the frequency and amplitude of sEPSCs recorded from pyramidal neurons in the CA1 region of the hippocampus in 1-month-old WT (**D**) and Pyk2 KO (**F**) mice treated or not with 100 nM Aβo for 20 min (D: $n = 9$ neurons per condition from 3 WT mice with 3 neurons per mouse; F: $n = 13$ neurons per condition from 4 Pyk2 KO mice with 3 to 4 neurons per mouse). Frequency and amplitude are expressed as the ratio of the value at 20 min (T20) to the value at the beginning of the recording (T0). Mann–Whitney tests, ns = not significant

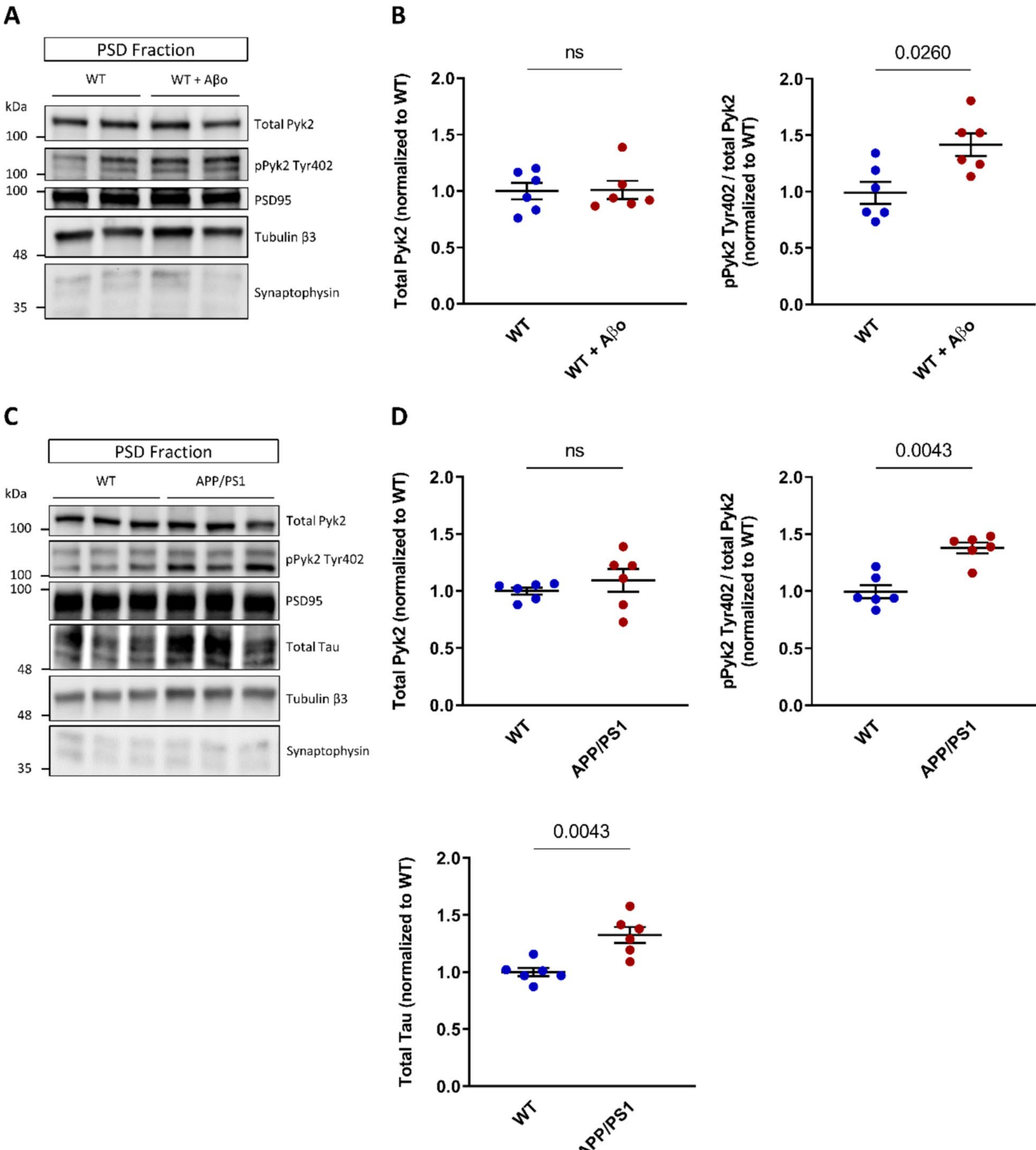

**Fig. 2** During neuronal hyperactivity, Pyk2 is phosphorylated on its tyrosine 402. **A, C** Representative immunoblots of hippocampal PSD fractions from 1-month-old WT mouse slices treated or not with 100 nM Aβo for 20 min (**A**), and from 1-month-old WT and APP/PS1 mouse slices (**C**). The enrichment of PSD95 and the absence of synaptophysin in the postsynaptic fraction confirm the quality of the fractionation. **B, D** Quantification of immunoblots as in (**A**) and (**C**), respectively. Total Pyk2, pPyk2 Tyr402/total Pyk2 and total Tau were normalized to tubulin β3. Data are plotted as mean ± SEM (n = 6 mice per group). Mann–Whitney tests, ns = not significant

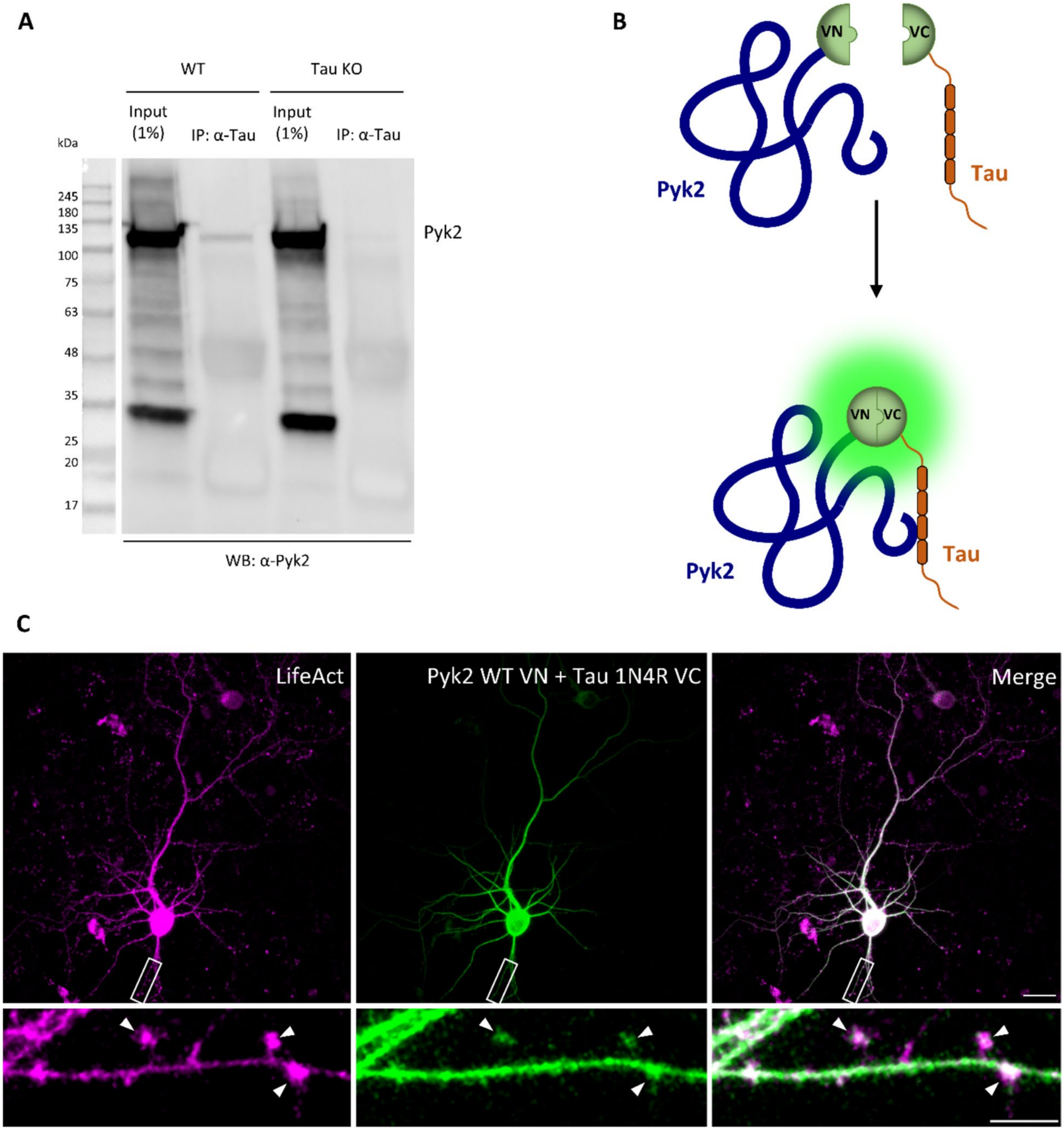

**Fig. 3** Pyk2 interacts with Tau in neurons. **A** Tau was immunoprecipitated from cortical lysates using anti-total Tau antibody. Immunoblots were subsequently performed to detect Pyk2 co-immunoprecipitation ($n=2$). The amount of input loaded was 1% of the protein from the total lysate used for immunoprecipitation. Tau KO mice were used as a negative control. **B** Schematic representation of the bimolecular fluorescence complementation technique: Pyk2 is tagged with a half-GFP called VN, while Tau is tagged with the complementary half-GFP called VC. If these two proteins interact, the two GFP halves complement each other, resulting in green fluorescence emission. **C** Representative confocal images of cultured cortical neurons overexpressing LifeAct-RFP (in magenta) and Pyk2 WT VN + Tau 1N4R VC (in green). Scale bar = 20 µm. Insets are magnifications of the white rectangle regions (scale bar = 5 µm). White arrows highlight dendritic spines exhibiting Pyk2–Tau interaction

demonstrating the specificity of the Pyk2– Tau interaction in neurons.

### Pyk2 contributes to Tau synaptic localization

In Alzheimer's disease, Aβo have been shown to cause the translocation of Tau to dendritic spines, leading to synaptotoxic effects [43]. Here, we demonstrated that Aβo induce Pyk2 activation in the PSD fraction of hippocampal neurons (see Fig. 2A and B). Furthermore, BiFC experiments revealed an interaction between Pyk2 and Tau in dendritic spines (see Fig. 3). Consequently, we investigated the impact of Pyk2 expression on Tau location in dendritic spines.

First, we explored the effect of Pyk2 absence on Tau expression. We carried out immunoblots on hippocampal slice lysates from 3-month-old WT and Pyk2 KO mice (Fig. 4A and B). In total lysates, no modification in Tau expression was observed in Pyk2 KO mice compared to WT mice. We then focused on the impact of Pyk2 deletion at the synaptic level. We performed cellular fractionation on hippocampi from 3-month-old WT and Pyk2 KO mice (Fig. 4C and D). In the PSD fraction of Pyk2 KO mice, we observed a 34% decrease in total Tau expression compared to WT. These results suggest that Pyk2 contributes to the synaptic localization of Tau.

### Pyk2 is involved in APPswe overexpression-induced synaptic loss

We showed the involvement of Pyk2 in Aβo-induced hippocampal neuronal hyperactivity in the early stages of an AD mouse model. As Pyk2 is also involved in postsynaptic organization [26, 44], we hypothesized that Pyk2 promotes Aβo-induced synaptic loss observed in APP/PS1 mice. To test this hypothesis, we transfected primary cortical neurons derived from WT and Pyk2 KO mouse embryos with a plasmid encoding APPswe, which induces Aβ overproduction in neurons (Fig. 5). Overexpression of APPswe in WT neurons resulted in a 18% decrease in spine density. However, Pyk2 KO neurons showed no significant difference in spine density between control and neurons overexpressing APPswe. The absence of dendritic spine density reduction in Pyk2 KO neurons suggests a role for Pyk2 in Aβo-induced synaptotoxicity.

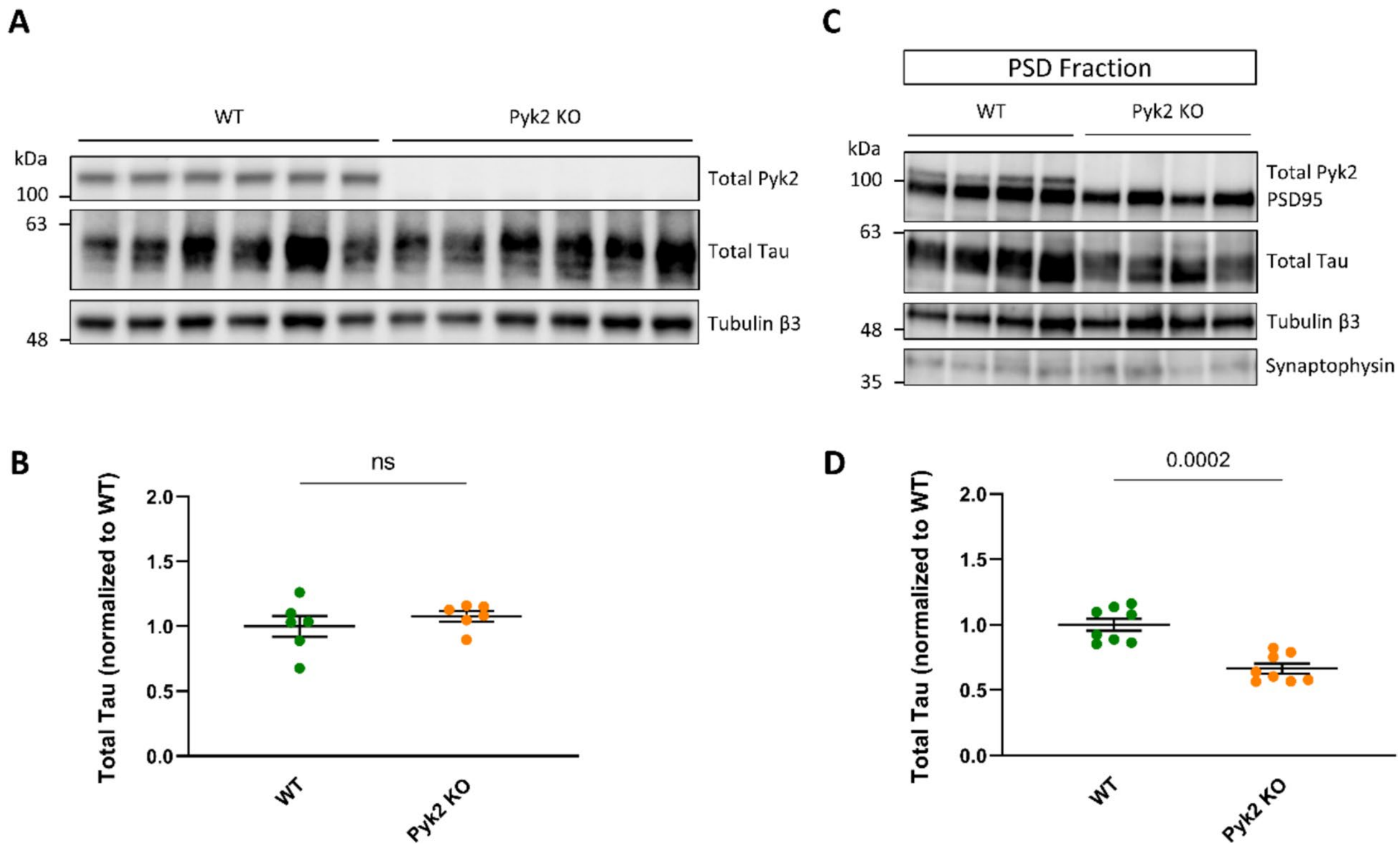

**Fig. 4** Pyk2 promotes synaptic localization of Tau. **A, C** Representative immunoblots of hippocampal total lysates (**A**) and PSD fractions (**C**) from 3-month-old WT and Pyk2 KO mice. In (**C**), the enrichment of PSD95 and the absence of synaptophysin in the postsynaptic fraction confirm the quality of the fractionation. **B, D** Quantification of immunoblots as in (**A**) and (**C**), respectively. Total Tau was normalized to tubulin β3. Data are plotted as mean ± SEM (n = 6 (**B**) and 8 (**D**) mice per group). Mann–Whitney tests, ns = not significant

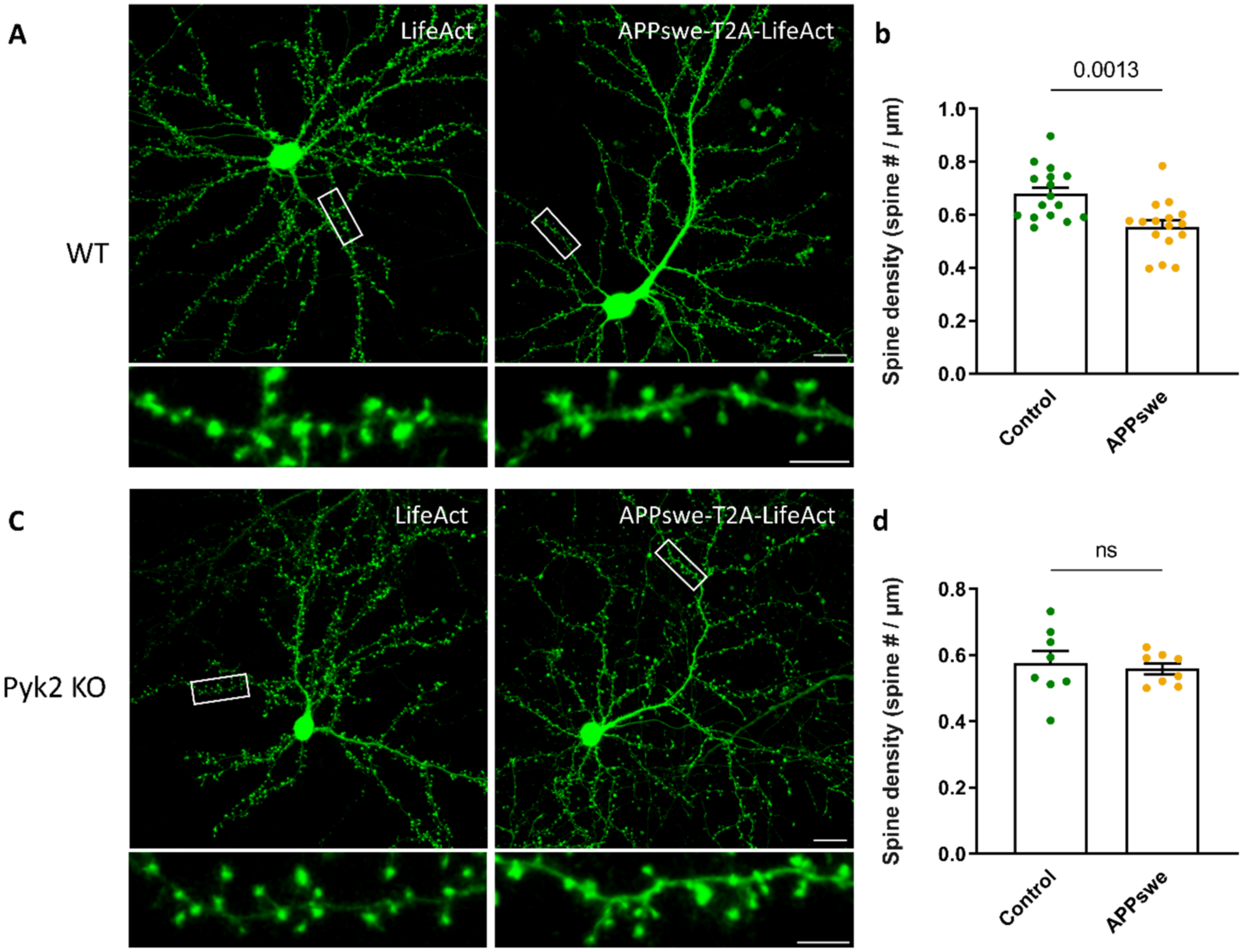

**Fig. 5** Genetic suppression of Pyk2 prevents APPswe-induced decrease in dendritic spine density. **A, C** Representative confocal images of cultured cortical neurons from C57BL/6J WT (**A**) and Pyk2 KO (**C**) mice overexpressing LifeAct-GFP or APPswe-GFP-T2A-LifeAct-GFP. Scale bar = 20 µm. Insets are magnifications of the white rectangle regions (scale bar = 5 µm). **B, D** Quantification of dendritic spine density of transfected neurons under the same conditions as in (**A**) and (**C**), respectively. Control and APPswe conditions correspond to neurons transfected solely with LifeAct-GFP or with APPswe-GFP-T2A-LifeAct-GFP, respectively. Data are plotted as mean ± SEM (**B**: Control, n = 16 and APPswe, n = 16 from at least 4 different cultures; **D**: Control, n = 8 and APPswe, n = 8 neurons from at least 2 different cultures). Mann–Whitney tests, ns = not significant

### Aβo trigger Pyk2 phosphorylation at Tyr402 specifically in dendritic spines

Given our observation of an Aβo-triggered increase in Pyk2 phosphorylation in hippocampal slices from 1-month-old APP/PS1 mice and Aβo-treated WT mice (see Fig. 2), we sought to validate this effect in our culture model. We did not detect any change in Pyk2 phosphorylation in total lysates of cultured neurons treated with 100 nM Aβo for 1, 6, or 24 h (Fig. 6A and B). We then performed cellular fractionation on cultured neurons exposed to the same treatment conditions. In the PSD fraction, we observed an approximately 40% increase in Pyk2 phosphorylation at Tyr402 after 1 h of treatment, which persisted up to 24 h, with no change in total Pyk2 expression (Fig. 6C and D). These results suggest that Aβo's effect on Pyk2 phosphorylation is specifically localized to the postsynaptic compartment of neurons.

### Pyk2 overexpression leads to a decrease in dendritic spine density independently of its activation or kinase activity, but through its proline-rich motif 1

We demonstrated that Pyk2 deletion prevented Aβo-induced synaptic loss in cultured cortical neurons. To understand Pyk2's role in synaptic density reduction, we transfected neurons with plasmids encoding mutated forms of Pyk2 tagged with a mCherry fluorophore.

Given that Aβo lead to Pyk2 phosphorylation, we first overexpressed Pyk2 WT and two mutated forms of Pyk2:

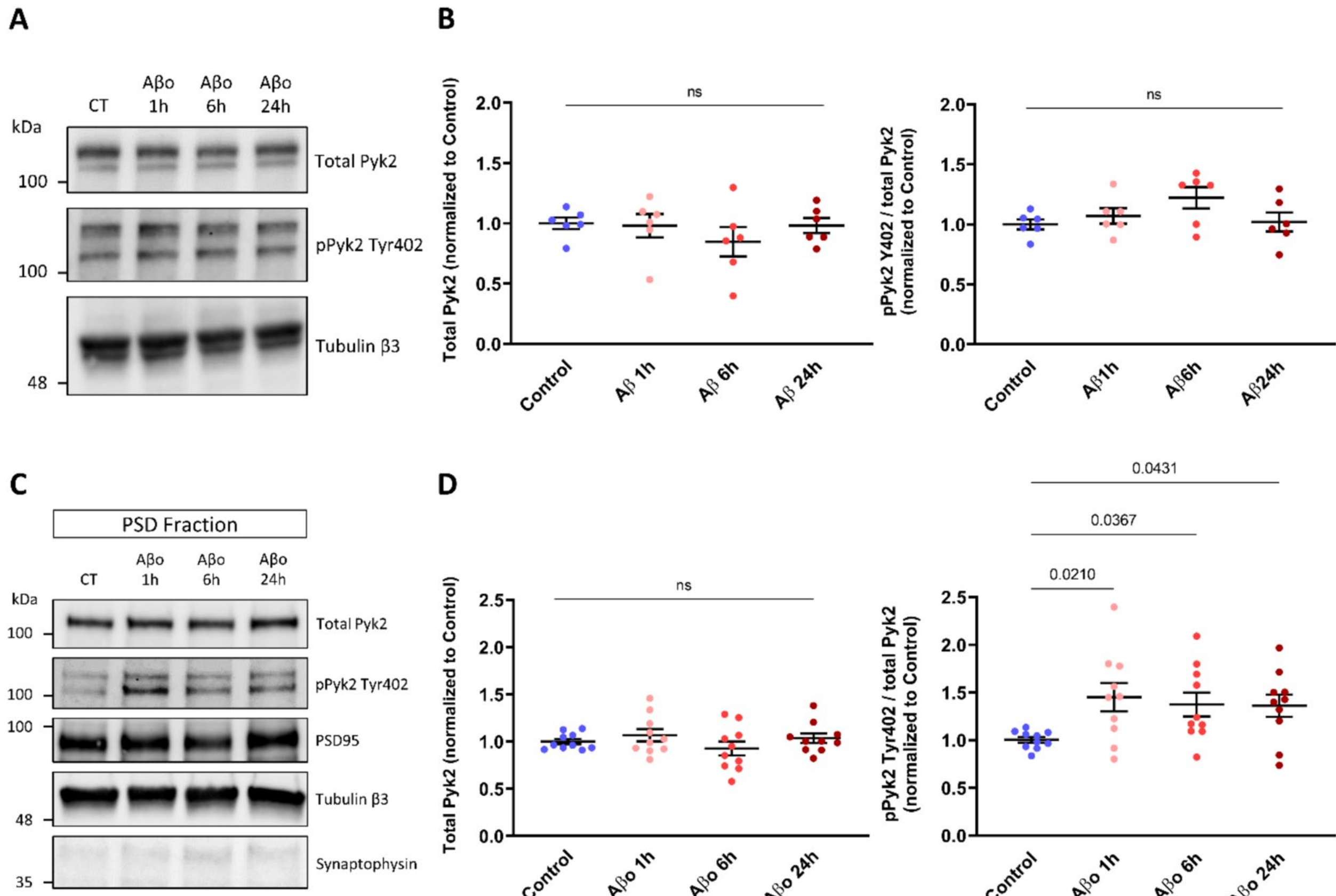

**Fig. 6** Aβo exposure induces increased phosphorylation of Pyk2 Tyr402 specifically in the postsynaptic compartment. **A, C** Representative immunoblots of total lysates (**A**) and PSD fractions (**C**) from cultured cortical neurons treated or not with 100 nM Aβo for 1, 6, or 24 h. In (**C**), the enrichment of PSD95 and the absence of synaptophysin in the postsynaptic fraction confirm the quality of the fractionation. **B, D** Quantification of immunoblots as in (**A**) and (**C**), respectively. Total Pyk2 and pPyk2 Tyr402/total Pyk2 were normalized to tubulin β3. Data are plotted as mean ± SEM (n = 6 (**B**) and 10 (**D**) different cultures per condition). Kruskal–Wallis tests followed by Dunn's post-hoc test, ns = not significant

**Fig. 7** Pyk2 overexpression in cultured neurons decreases spine density independently of its activation or kinase activity. **A** Confocal images of a cortical neuron from mouse embryos in culture. Neurons were transfected with LifeAct-GFP (in green) and mCherry-Pyk2 WT (in magenta) at 13 DIV and imaged 48 h after transfection. Scale bar = 20 µm. Insets are magnifications of the white rectangle regions (scale bar = 5 µm). White arrows highlight dendritic spines exhibiting Pyk2 expression. **B** Representative confocal images of cultured cortical neurons overexpressing either LifeAct-GFP alone or along with mCherry-Pyk2 WT, Y402F or K457A. Only images with LifeAct-GFP are shown. Scale bar = 20 µm. Insets are magnifications of the white rectangle regions (scale bar = 5 µm). **C** Representation of Pyk2 mutants used in (**B**). Circles indicate the positions of the mutations (Y402F and K457A) within the Pyk2 sequence. **D** Quantification of dendritic spine density of transfected neurons under the same conditions as in (**B**). The control condition refers to neurons transfected solely with LifeAct-GFP. Data are plotted as mean ± SEM (Control, n = 26; Pyk2 WT, n = 23; Pyk2 Y402F, n = 20; Pyk2 K457A, n = 24 neurons from at least 6 different cultures). Kruskal–Wallis test followed by Dunn's post-hoc test

one with a mutation at its trans-autophosphorylation site (Pyk2 Y402F) that prevents Pyk2 phosphorylation and therefore its activation, and the other with a mutation in its kinase domain (Pyk2 K457A) that abolishes its kinase activity (Fig. 7A-C). Overexpression of Pyk2 WT resulted in a ubiquitous distribution in the soma, neurites and dendritic spines (Fig. 7A). Although we illustrated only the Pyk2 WT distribution in Fig. 7, it should be noted that the Pyk2 mutants used in this study exhibited similar subcellular distributions despite variable expression levels in HEK293 cells (Supplementary Fig. 3). When neurons overexpressed Pyk2 WT, a 22% decrease in dendritic spine density was observed compared to control neurons (Fig. 7D). A similar decrease was found in neurons overexpressing Pyk2 Y402F or K457A (22% and 16% decrease, respectively). However, no significant difference was observed between neurons transfected with the plasmid encoding Pyk2 WT and those expressing either of its mutants. Hence, overexpression of Pyk2 results in a decrease in dendritic spine density, independent of its activation or kinase activity.

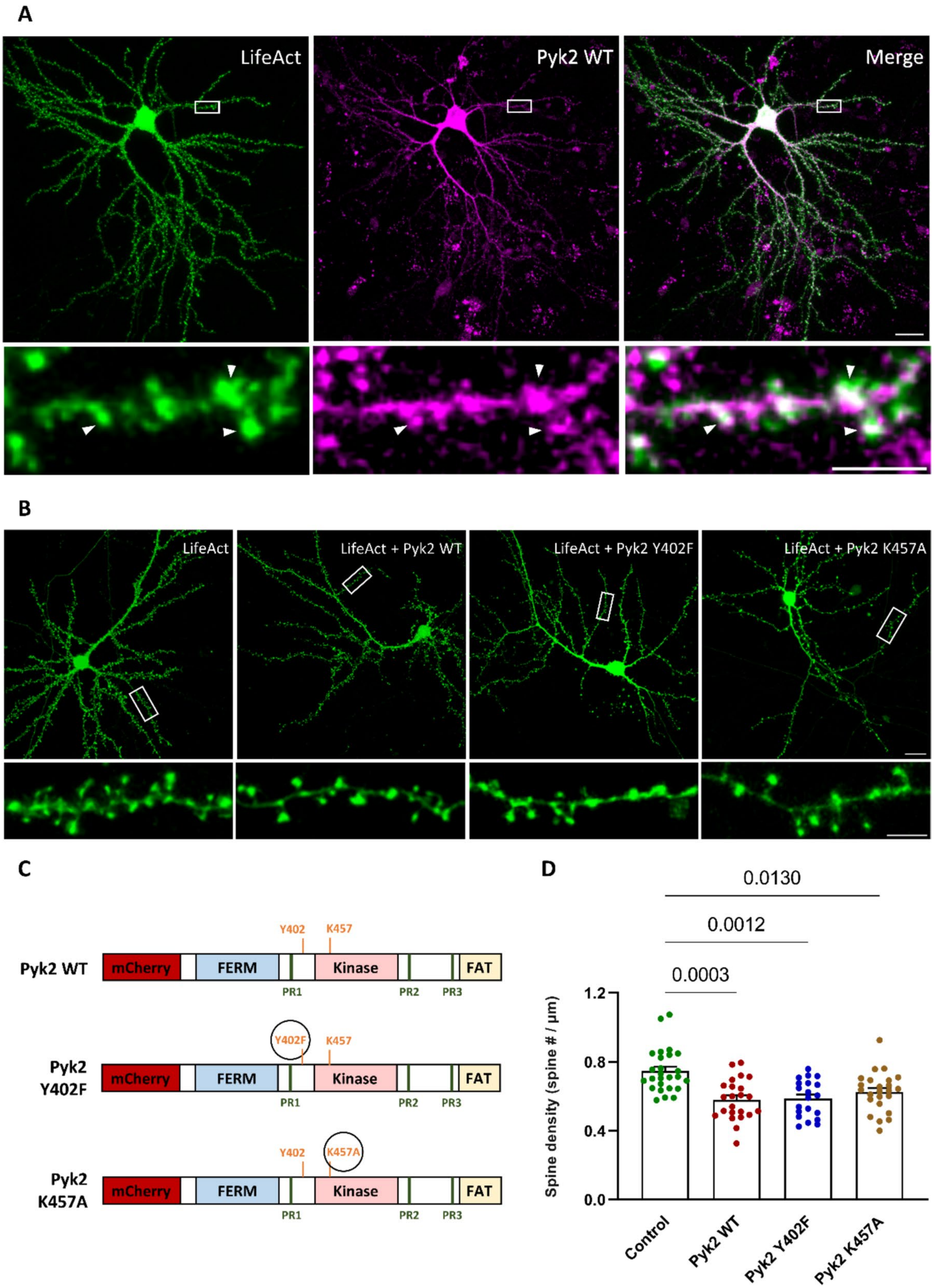

**Fig. 7** (See legend on previous page.)

We then focused on the distinct domains of Pyk2 that may be involved in the alteration of spine density. To address this question, we transfected truncated forms of Pyk2: one containing only the FERM domain ($Pyk2_{1-368}$) and the other lacking both the FERM domain and the FERM-kinase linker containing the PR1 motif and the autophosphorylation site Tyr402 ($Pyk2_{421-1009}$) **(**Fig. 8A-C**)**. Neurons overexpressing either of the truncated forms of Pyk2 exhibited an increase in dendritic spine density compared to those transfected with Pyk2 WT, reaching levels comparable to the control condition.

The two mutants that did not decrease spine density both lacked the FERM-kinase linker region. Since the autophosphorylation site mutant also induced a reduction in spine density, we focused on the PR1 motif, as this is the only functionally identified sequence missing from the two truncated forms that did not decrease spine density. We generated a plasmid encoding a Pyk2 variant in which proline residues 413 and 416 within the PR1 motif were replaced with alanine (Pyk2 P413/416A) **(**Fig. 8D-F**)**. Neurons overexpressing the PR1 mutant of Pyk2 exhibited significantly higher spine density compared to those transfected with Pyk2 WT. This density was similar to that observed in control LifeAct-GFP transfected neurons. Taken together, these data demonstrate that overexpression of Pyk2 leads to a decrease in dendritic spine density, independent of its activation or kinase activity, but through a molecular mechanism involving the PR1 motif.

## Discussion

This study aimed to better understand the role of Pyk2 in the early pathophysiological alterations of AD. We demonstrated that Pyk2 is involved in the two major early features of the disease: hippocampal neuronal hyperactivity and synaptic loss. Hippocampal hyperactivity has been reported in the prodromal stage of AD, while cortical hyperactivity has been observed in the preclinical stage [19, 22]. This Aβo-induced neurotransmission impairment has also been recorded in the hippocampus of the APP/PS1-21 mouse model of AD and in WT mice acutely exposed to Aβo [24, 37]. Our team has previously shown that inhibiting hippocampal neuronal hyperactivity in APP/PS1 mice prevents synaptic loss and cognitive deficits that occur later in disease progression [24]. Here, we show that genetic deletion of Pyk2 prevents Aβo-induced hippocampal neuronal hyperactivity, implicating Pyk2 in this early synaptic dysfunction. A causal link between Pyk2 and Aβo-driven disruption of excitatory neurotransmission during synaptic plasticity has previously been reported [11], further supporting its key role in AD-related neurotransmission impairment. These findings suggest that early-stage inhibition of Pyk2 expression, before the onset of structural or cognitive alterations, could potentially lead to early therapeutic interventions for AD. Notably, hippocampal neuronal hyperactivity in APP/PS1 mice and Aβo-treated WT mice correlates with increased phosphorylated Pyk2 in the postsynaptic compartment. Given that Pyk2 is a calcium-dependent tyrosine kinase [40, 45], its activation through autophosphorylation at tyrosine 402 may result from the increased neuronal calcium influx during neuronal hyperactivity [39]. Although our data indicate that genetic deletion of Pyk2 prevents Aβo-induced hippocampal hyperactivity, it remains to be elucidated whether Pyk2 expression per se, its autophosphorylation, and/or its kinase activity mediates this effect.

In AD, synaptic loss is a critical factor leading to cognitive decline [46–50]. We demonstrate that genetic deletion of Pyk2 protects against Aβo-induced synaptic loss in cultured cortical neurons. We also observe that Pyk2 overexpression induces synaptic loss. Consistent with this observation, previous studies have shown that Pyk2 overexpression decreases synaptic density in cultured hippocampal neurons [10]. In our model, Pyk2's involvement in synaptic loss is independent of its activation and kinase activity but depends on its PR1 motif, highlighting a potential pathological scaffolding role for Pyk2 in synaptotoxicity. Indeed, proline-rich motifs are often involved in protein–protein interactions. The Pyk2 PR1 motif is likely to be a binding site for the Src SH3 domain, as shown for the similar motif in FAK [51]. Whether the recruitment of Src-family kinases – including Fyn, a known Tau kinase [14, 52] – or other partners to the PR1 motif contributes to the neurotoxic effects of Pyk2 will require further investigation. Nevertheless, our data suggest that, in addition to its kinase function, Pyk2 exhibits scaffolding properties that contribute to Aβo-driven synaptotoxicity. This scaffolding role is further supported by recent findings showing that Pyk2 suppresses contextual fear memory, independent of its activation, in Pyk2 Y402F mutant mice [53]. While our results reveal that the Pyk2 PR1 motif is involved in synaptic loss, a potential role for calcium cannot be excluded. Pyk2 is natively folded in a closed conformation due to an intramolecular interaction between its FERM and kinase domains. Binding of the calcium/calmodulin complex to specific sites on Pyk2 promotes the release of this interaction, thereby exposing potential binding sites, such as the PR1 motif, for Pyk2 partners [54, 55].

**Fig. 8** Pyk2 overexpression leads to a decrease in dendritic spine density through its PR1 motif. **A** Representative confocal images of cultured cortical neurons overexpressing LifeAct-GFP and mCherry-Pyk2 WT, (1–368) or (421–1009). Only images with LifeAct-GFP are shown. Scale bar = 20 µm. Insets are magnifications of the white rectangle regions (scale bar = 5 µm). **B** Representation of Pyk2 truncated forms used in (**A**). **C** Quantification of dendritic spine density of transfected neurons under the same conditions as in (**A**). The control condition corresponds to neurons transfected solely with LifeAct-GFP. Data are plotted as mean ± SEM (Control, n = 26; Pyk2 WT, n = 24; $Pyk2_{1-368}$, n = 23; $Pyk2_{421-1009}$, n = 23 neurons from at least 6 different cultures). Kruskal–Wallis test followed by Dunn's post-hoc test. **D** Representative confocal images of cultured cortical neurons overexpressing LifeAct-GFP and mCherry-Pyk2 WT or Pyk2 P413/416A. Only images with LifeAct-GFP are shown. Scale bar = 20 µm. Insets are magnifications of the white rectangle regions (scale bar = 5 µm). **E)** Representation of Pyk2 mutant used in (**D**). The encircled part indicates the position of the P413/416A mutation within the Pyk2 sequence. **F)** Quantification of dendritic spine density of transfected neurons under the same conditions as in (**D**). The control condition corresponds to neurons transfected solely with LifeAct-GFP. Data are plotted as mean ± SEM (Control, n = 18; Pyk2 WT, n = 16; Pyk2 P413/416A, n = 16 neurons from at least 5 different cultures). One-way ANOVA test followed by Tukey's post-hoc test

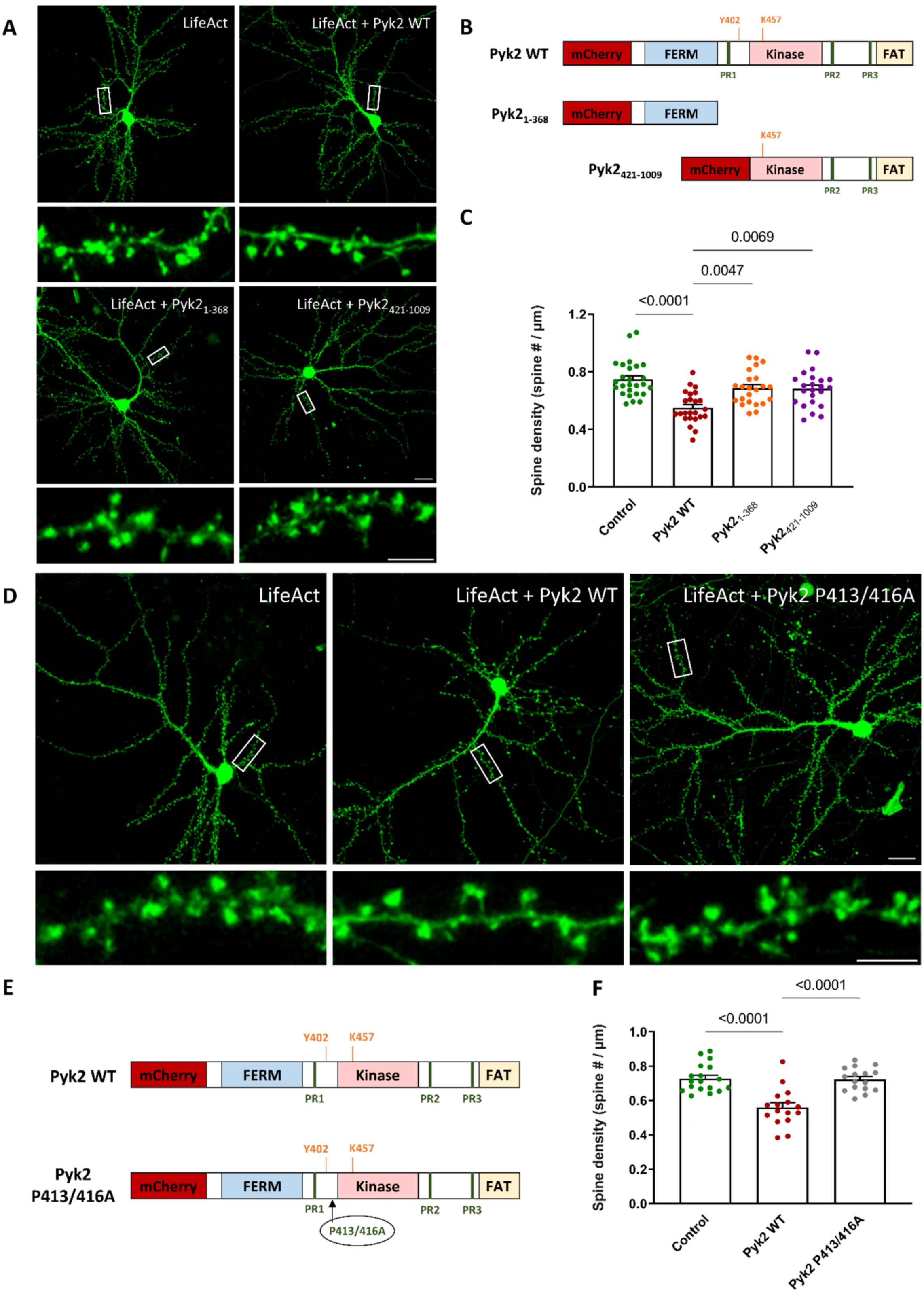

**Fig. 8** (See legend on previous page.)

We demonstrated the involvement of Pyk2 in the pathophysiological alterations occurring in the early stages of AD. These results are consistent with previous studies showing that genetic deletion of Pyk2 in APPswe/PS1ΔE9 mice prevents the onset of cognitive deficits [11]. Conversely, another study reported that overexpression of Pyk2 in the hippocampus of 8-month-old 5xFAD mice increases synaptic density and improves behavioral impairments [12]. These contrasting results suggest a dual role for Pyk2 depending on the stage of the disease, emphasizing the critical need for further investigation to fully elucidate its mechanisms of action throughout Alzheimer's disease progression.

Among the protein partners of Pyk2, we show that Pyk2 interacts with Tau in neurons, supporting previous reports describing Pyk2–Tau interactions both *in vitro* and *in vivo* [14]. Our bimolecular fluorescence complementation assays confirmed Pyk2–Tau interactions in various neuronal compartments, including soma, dendrites, and synapses. Moreover, we found that genetic deletion of Pyk2 did not alter total Tau expression in hippocampal lysates but reduced its expression in the postsynaptic fraction, highlighting a compartment-specific effect of Pyk2. Thus, while Pyk2 appears to promote the synaptic localization of Tau, the precise contribution of its expression, activation, or kinase activity remains unclear. In addition, whether Tau contributes to the Pyk2-mediated decrease in synaptic density remains to be investigated. Recent studies have highlighted that the abnormal presence of Tau oligomers in synapses of patients with dementia may act as a signal for glial cells to eliminate these synapses, leading to synaptic loss and cognitive deficits [56]. Although the underlying mechanisms by which Pyk2 affects Tau synaptic localization are not yet fully understood, this process may contribute to the progression of synaptic dysfunction and cognitive impairments.

## Conclusion

Pyk2, a tyrosine kinase predominantly expressed in neurons, has been identified as a genetic risk factor for late-onset Alzheimer's disease. Our study shows that Pyk2 is involved in early AD-related alterations induced by Aβ oligomers. We demonstrate, for the first time, the implication of Pyk2 in hippocampal neuronal hyperactivity, the initial pathological event in AD that precedes structural or cognitive impairments. Moreover, we show that beyond its kinase activity, Pyk2 exhibits crucial scaffolding properties that contribute to Aβo-induced synaptic loss.

Furthermore, we establish that Pyk2 promotes Tau mislocalization to synapses, a phenomenon known to contribute to synaptic dysfunction in AD. This effect is specific to the postsynaptic compartment, where Pyk2 is activated during neuronal hyperactivity.

While the precise mechanisms by which Pyk2 influences AD development remain to be fully elucidated, our data provide compelling evidence for Pyk2 as a promising early therapeutic target for synaptotoxicity in Alzheimer's disease. These findings not only advance our understanding of Pyk2's complex role in AD pathogenesis but also provide novel insights for preventive strategies in this neurodegenerative disorder.

### Abbreviations

Aβ Amyloid-beta
Aβo Amyloid-beta oligomers
aCSF Artificial cerebrospinal fluid
AD Alzheimer's disease
ANOVA Analysis of variance
APPswe Human amyloid precursor protein carrying the Swedish mutation
APP/PS1 APP/PS1-21
BACE1 Beta-site APP cleaving enzyme 1
BCA Bicinchoninic acid
BiFC Bimolecular fluorescence complementation
BSA Bovine serum albumin
DIV Days in vitro
DMEM Dulbecco's Modified Eagle's Medium
EDTA Ethylenediaminetetraacetic acid
EGTA Ethyleneglycol- bis(β-aminoethyl)-N,N,N′,N′-tetraacetic acid
GSK-3β Glycogen synthase kinase 3 beta
HFIP 1,1,1,3,3,3-Hexafluoro-2-propanol
HRP Horseradish peroxidase
KO Knock-out
MCI Mild cognitive impairment
Ns Not significant
PBS Phosphate-buffered saline

| | |
|---|---|
| PR1 | Proline-rich motif 1 |
| PS1 | Presenilin-1 |
| PSD | Postsynaptic density |
| PSD95 | Postsynaptic density protein 95 |
| PTK2B | Protein tyrosine kinase 2 beta |
| Pyk2 | Proline-rich tyrosine kinase 2 |
| RIPA | Radioimmunoprecipitation assay buffer |
| SDS-PAGE | Sodium dodecyl sulfate–polyacrylamide gel electrophoresis |
| SEM | Standard error of the mean |
| sEPSCs | Spontaneous excitatory postsynaptic currents |
| TBS | Tris-buffered saline |
| TBS-T | Tris-buffered saline with Tween 20 |
| WT | Wild-type |

## Supplementary Information

The online version contains supplementary material available at https://doi.org/10.1186/s44477-026-00021-4.

Below is the link to the electronic supplementary material.Supplementary file1 (DOCX 4425 KB)

Supplementary file2 (PDF 904 KB)

**Acknowledgements**

We thank Dr. Isabelle Arnal's team at the Grenoble Institute of Neurosciences for providing Tau KO mice.

**Author's contributions**

Q.R. and A.B. conceptualized and designed the study. Q.R. performed major experiments and data analysis. F.P. conducted and analysed patch clamp experiments. K.V.B., E.B. and S.B. carried out experiments. B.B. and J.A.G. provided expertise and materials. A.B. supervised the study. Q.R. wrote the initial draft, and all authors participated in its editing. All authors read and approved the final manuscript.

**Funding**

This work was supported by INSERM, University Grenoble Alpes, and the French National Research Agency (ANR-19-CE16-0020).

**Data availability**

All data generated or analyzed during this study are included in the article and its supplementary information file. Requests for raw data should be addressed to the corresponding authors upon reasonable request.

## Declarations

**Ethics approval and consent to participate**

All experiments were conducted in accordance with the European Community Council directive 86/609/EEC (November 24, 1986) and French national institutional animal care guidelines (protocol APAFIS#45114). All experimental protocols were approved by the Grenoble Institute of Neurosciences Ethics Committee.

**Consent for publication**

Not applicable.

**Competing interests**

The authors declare no competing interests.

**Author details**

[1]Univ. Grenoble Alpes, Inserm, U1216, Grenoble Institut Neurosciences, 38000 Grenoble, France. [2]Institut du Fer à Moulin, INSERM, Sorbonne University, Paris, France. [3]Paris Brain Institute, INSERM, Sorbonne University, Paris, France.

## Publisher's Note